\documentstyle[11pt,aaspp4]{article}

\def\grb{GRB~991216}

\received{2000 February 17}
\accepted{}
\journalid{}{}
\articleid{}{}

\slugcomment{To appear in {\it The Astrophysical Journal}}

\lefthead{Halpern et al.}
\righthead{Optical Afterglow of GRB 991216}

\begin{document}

\title{GRB 991216 Joins the Jet Set:\\ Discovery and Monitoring of its
Optical Afterglow}

\author{
J.~P.~Halpern\altaffilmark{1},
R.~Uglesich\altaffilmark{1},
N.~Mirabal\altaffilmark{1},
S.~Kassin\altaffilmark{2},
J.~Thorstensen\altaffilmark{3},
W.~C.~Keel\altaffilmark{4},
A.~Diercks\altaffilmark{5},
J.~S. Bloom\altaffilmark{5},
F.~Harrison\altaffilmark{5},
J.~Mattox\altaffilmark{6},
M.~Eracleous\altaffilmark{7}
}
\affil{}

\altaffiltext{1}{Astronomy Department, Columbia University, 550 West 120th Street, New York, NY 10027.}
\altaffiltext{2}{Department of Astronomy, Ohio State University,
140 West 18th Avenue, Columbus, OH 43210.} 
\altaffiltext{3}{Department of Physics \& Astronomy, Dartmouth College,
6127 Wilder Hall, Hanover, NH 03755.}
\altaffiltext{4}{Department of Physics \& Astronomy, University of Alabama,
Box 870324, Tuscaloosa, AL 35487.} 
\altaffiltext{5}{Palomar Observatory 105-24, California Institute of
Technology, Pasadena, CA 91125.}
\altaffiltext{6}{Department of Astronomy, Boston University,
725 Commonwealth Ave., Boston, MA 02215.} 
\altaffiltext{7}{Department of Astronomy and Astrophysics,
The Pennsylvania State University, 525 Davey Laboratory, University Park,
PA 16802.}

\begin{abstract}
\rightskip 0pt \pretolerance=100 \noindent
The optical light curve of the energetic $\gamma$-ray
burst \grb\ is consistent with jet-like behavior in which a power-law
decay steepens from $t^{-1.22 \pm 0.04}$ at early times to
$t^{-1.53 \pm 0.05}$ in a gradual transition at around 2~d.
The derivation of the late-time decay slope takes into
account the constant contribution of a host or intervening galaxy
which was measured 110 days after the event at $R = 24.56 \pm 0.14$,
although the light curve deviates from a
single power law whether or not a constant term is included.
The early-time spectral energy distribution of the afterglow
can be described as $F_{\nu} \propto \nu^{-0.74 \pm 0.05}$ 
or flatter between
optical and X-ray, which, together with the slow initial decay,
is characteristic of standard adiabatic evolution in a uniformly
dense medium.   Assuming that a reported absorption-line
redshift of 1.02 is correct,
the apparent isotropic energy of $6.7 \times 10^{53}$~ergs is reduced
by a factor of $\approx 200$ in the jet model, and the
initial half-opening angle is $\approx 6^{\circ}$.  \grb\
is the third good example
of a jet-like afterglow (following GRB~990123 and GRB~990510),
supporting a trend in which the apparently
most energetic $\gamma$-ray events have the narrowest collimation
{\it and} a uniform ISM environment.  This, plus the absence of evidence
for supernovae associated with jet-like afterglows, suggests that
these events may originate from a progenitor
in which angular momentum plays an important role but a
massive stellar envelope or wind does not, {\it e.g.},
the coalescence of a compact binary.

\end{abstract}

\section{Introduction}

\grb\ was one of the brightest $\gamma$-ray bursts detected by
the Burst and Transient Source Experiment (BATSE),
with a fluence of $(2.56 \pm 0.01) \times 10^{-4}$ ergs~cm$^{-2}$
above 20.6~keV (Kippen 1999).  Rapid follow-up by the
{\it Rossi X-ray Timing Explorer\/} Proportional Counter Array
({\it RXTE\/} PCA) detected a fading X-ray afterglow in two sets of
scans 4.0 and 10.9~hr after the burst (Takeshima et~al. 1999).
We began optical observations at 10.8~hr using the 1.3m telescope
of the MDM Observatory, covering a $17^{\prime}\times 17^{\prime}$
field that was large enough to encompass both preliminary and
final {\it RXTE\/} derived positions.  We discovered the
optical afterglow (Uglesich et~al. 1999), initially at $R = 18.5$,
during the course of 8~hr of nearly continuous monitoring,
by employing the image subtraction
technique of Tomaney \& Crotts (1996) to search the entire field
at once for variable objects.
Its position is (J2000) $05^{\rm h}09^{\rm m}31.\!^{\rm s}29,
+11^{\circ}17^{\prime}07.\!^{\prime\prime}4$ in the
USNO--A2.0 reference system (Monet et al. 1996).

Radio observations within the first two days detected a compact,
variable source with flux density $\sim 1$~mJy at 4.8, 8.5 and 15~GHz
(Taylor \& Berger 1999; Rol et al. 1999; Pooley 1999).
A VLBA observation
40~hr after the burst yielded a size less than 1~mas
(Taylor \& Frail 1999).  An spectrum of the optical transient (OT)
obtained on the VLT-Antu telescope revealed three systems of absorption lines,
at $z=0.77$, $z=0.80$, and $z=1.02$ (Vreeswijk et al. 1999).
If the burst were located in the highest redshift system, its 
isotropic $\gamma$-ray energy was $6.7 \times 10^{53}$~ergs
(assuming $H_0 = 65$ km~s$^{-1}$~Mpc$^{-1}$,
$\Omega_0 = 0.2$, $\Lambda = 0$), second only to GRB~990123
($3.4 \times 10^{54}$~ergs, Kulkarni et al. 1999)
among those bursts with measured redshifts.

A key question about energetic bursts such as
this one is whether evidence for collimation can be found,
and the extent to which such jet behavior reduces the
total inferred energy to a value compatible with
that available in compact-object binaries. The principal manifestation
of a jet geometry is a gradual achromatic steepening of the light curve to
a $\approx t^{-2}$ decay after the edge of the jet becomes visible
and shortly thereafter begins to spread laterally
(Panaitescu, M\'esz\'aros, \& Rees 1998;
Rhoads 1999; Sari, Piran, \& Halpern 1999;
Moderski, Sikora, \& Bulik 2000; Kumar \& Panaitescu 2000).
Alternatively, such a steepening may occur when the jet
becomes non-relativistic (Huang, Dai, \& Lu 1999,2000).
In this paper, we present evidence for a jet in \grb.

Also of interest is any information that can be gleaned about the environment
in which the burst occurs, and what this suggests about the
possible progenitor star(s).  The spectral
and temporal evolution of the afterglow can
distinguish between a uniform interstellar medium (ISM), and
a nonuniform medium of density $n \propto r^{-2}$, e.g., as is
appropriate for a pre-existing stellar wind (Chevalier \& Li 1999,2000).
We will argue that the afterglow observations of \grb\ are more
consistent with a uniform ISM.

\section{Optical and Infrared Observations}

Optical and IR photometry of \grb\ was collected at various telescopes
as listed in Table~1. 
We placed all of the optical observations
on a common $VR_cI_c$ system using the calibrations made of this
field by Henden, Guetter, \& Vrba (1999).  In this system, stars
A and B in Figure~1 have $R_{\rm c} = 15.345$
and $R_{\rm c} = 19.478$, respectively.
Except for the final point on January 13, all of the MDM data
were obtained through non-standard broad $R$ and $I$
filters which we calibrated using 66 Landolt (1992) standard stars.
The rms scatter in the fitted transformation is 0.024 mag,
which we have included in Table~1 as a systematic error.
Gunn $r$ and $i$ measurements at Palomar and Keck were also
transformed to $R_{\rm c}$ and $I_{\rm c}$ using Landolt standards.
For the $JHK$ observations at MDM, we used UKIRT
faint standard stars for calibration.  In Figure~2 we graph the
light curves in $R$, $I$, and $J$, together with a model developed
for the $R$ band as described below.

Beginning 10.8~hr after the burst, we obtained a nearly continuous
set of $R$-band observations for 8~hr on the MDM 1.3m telescope.
This uniform set of data, shown in Figure~3,
can be fitted assuming statistical uncertainties
only with a power-law decay of
$\alpha = -1.219 \pm 0.036$.  The rms scatter about this fit is only 
0.019 mag, which is less than the typical statistical uncertainty
of 0.03 mag in the individual points.  Therefore, we conclude that
any intrinsic fluctuation about the mean decay
on time scales from 10 minutes to hours
is less than 1\%.  However, an extrapolation of this power law is
strongly inconsistent with the observations after about 3~d.
By day~10, when the extrapolation would predict $R = 22.66 \pm 0.11$,
we observe $R = 23.28 \pm 0.07$.  Furthermore, there is evidence that
the decay flattened after day~20, which we interpret as the increasing
dominance of a host or intervening galaxy.
In view of this behavior, we subtracted a constant flux
from each of the data points.  The value of this constant
was varied in the range $24.7 < R_0 < 24.9$; the measurement
of this contribution is explained in the following paragraph.
After day~1, all of the corrected data points are well fitted by
a steeper decay, $\alpha = -1.53 \pm 0.05$.  The error quoted here
includes both the statistical uncertainty in the fit ($\pm 0.02$)
and the the systematic range ($\pm 0.03$) introduced by the estimated
uncertainty in the subtracted galaxy contribution.

The fit of the late-time slope is dependent upon the interpretation
of the latest photometric point as dominated by the constant contribution
of a galaxy, which could be either the host or an intervening
galaxy in view of the three absorption-line redshifts seen
in the optical spectrum of the OT.
Here we describe some details of the
images that are relevant to this conclusion.
All of the images obtained after day~10, when we surmise that
galaxy contribution became significant, show some
extension to the west of the OT, and the penultimate image from January 13,
taken in $1.\!^{\prime\prime}0$ seeing, looks quite extended.
The final image obtained on 2000 April~3 under similar
seeing shows no indication of the OT.  Rather, an object extended by
$\approx 2.\!^{\prime\prime}5$ in the east-west direction
is centered approximately $\approx 0.\!^{\prime\prime}5$
west of the previously measured position of the OT.
In addition, there is a faint galaxy only $2.\!^{\prime\prime}3$
to the southeast of the OT, which makes it tricky to measure
an integrated magnitude for the OT/galaxy at late times.
The estimated total magnitude of the ``host'' galaxy is $R = 24.56 \pm 0.14$,
but not all of this flux is included in the photometry of the OT.
Instead, a lesser contamination of the OT is achieved by measuring its
magnitude after day~10 in a $1.\!^{\prime\prime}1$ radius aperture 
centered on the OT position.  This
procedure produces a relatively robust measurement of
the OT plus superposed galaxy contribution.
The galaxy contribution alone {\it at the position of the OT\/}
is measured in the same way on April~3.  The result is
$R = 24.8 \pm 0.1$, which is the final $R$ magnitude listed in Table~1
and the constant that we subtracted from the previous measurements
to fit the OT decay curve.

It has become customary to fit such steepening afterglow decays
with a smooth function that has asymptotic power-law behavior
at early and late times.  One such simple function is
$$ F(t)\ =\ {2\,F_*\,(t/t_*)^{\alpha_1} \over 
1+(t/t_*)^{(\alpha_1-\alpha_2)}}\ +\ F_0 \eqno(1) $$
Here $\alpha_1$ and $\alpha_2$ represent the 
asymptotic early and late time
slopes, $F_0$ is the constant galaxy contribution, 
and $F_*$ is the OT flux at the cross-over time $t_*$.
Unfortunately, the data on GRB~991216 do not permit many interesting
constraints on these parameters to be derived.
Our experiments with such fits show that $\alpha_2$ could be as steep as --2.1.  Therefore, we consider that $-2.1 < \alpha_2 < -1.5$ is 
the allowed range on the late-time decay, while $\alpha_1 \geq -1.22$ is certainly
true at early times.  There aren't observations
at sufficiently early times to constrain an upper limit on $\alpha_1$,
and $t_*$ is essentially unconstrained within a factor of 10
due to the sparse data at late times.
In Figure~2, we draw an acceptable fit to Equation~1
in which $\alpha_1 = -1.0$ was fixed, 
which results in best-fitted parameter values $\alpha_2 = -1.8$,
$t_* = 1.2$~d, and $R_0 = 24.76$.  The latter is 
consistent with $R_0 = 24.8 \pm 0.1$ as
the constant contribution of the portion of the
host or intervening galaxy which contaminates the OT photometry.
We stress that such a fit is illustrative only.  More important,
we note that the decay parameters of \grb\ are strikingly similar to
that of GRB~990123 (Kulkarni et al 1999),
the most energetic event yet observed.

\section{Broad-Band Continuum Shape and Reddening}

It is possible to synthesize a broad-band spectrum from these data by 
interpolating the magnitudes to a particular time using the observed
decay rates.
We chose a time of December 18.34 UT, 40~hr after the burst,
which falls close to the largest number of measurements at
different frequencies from radio through X-ray.
The interpolated $VRIJHK$ magnitudes were converted to fluxes using
standard calibrations, and are
graphed as filled circles in Figure~4.  Galactic
reddening is a significant factor in this field because of its
intermediate Galactic latitude, $(\ell,b) = (190.\!^{\circ}418,
-16.\!^{\circ}666)$.
The selective extinction $E(B-V)$ can be estimated in at least two
ways.  First is the value of Schlegel, Finkbeiner, \& Davis (1998)
from the {\it IRAS\/} $100\,\mu$m maps, $E(B-V) = 0.63$~mag.
This is significantly larger than
a second estimate, $E(B-V) = 0.40$~mag, which can be
derived from the Galactic 21 cm column density in this
direction, $N_{\rm HI} = 2.0 \times 10^{21}$~cm$^{-2}$
(Zhang \& Green 1991; Stark et al. 1992), and the standard
conversion $N_{\rm HI}/E(B-V) = 5.0 \times 10^{21}$~cm$^{-2}$~mag$^{-1}$
(Savage \& Mathis 1979).  It is likely that H~I underestimates
the extinction to \grb\ because this position falls on
the edge of a CO cloud that is part of an expanding molecular
and dust ring energized by the $\lambda$~Orionis H~II region $7^{\circ}$
away (Maddalena \& Morris 1987; Zhang et al. 1989). 
Associated with this CO cloud is the Lynds (1962) dark nebula
LDN~1571, only $0.\!^{\circ}7$ from \grb.

Figure~4 shows the results of applying
each of these extinction corrections.
In neither case is the dereddened spectrum a good fit
to a power law, although illustrative extreme fits are drawn
corresponding to the two suggested values of the extinction.
In particular, the turn-down in the $K$ band is puzzling
in view of the higher radio flux which was
observed at $\approx 1$~mJy on several occasions before and up to this time
(Taylor \& Berger 1999; Rol et al. 1999; Pooley 1999; Taylor \& Frail 1999).
Such a break would require a concave upward inflection at longer wavelengths,
which is not accommodated by any afterglow model.  A similar peak was
seen in the IR photometry of GRB~971214 (Ramaprakash et al. 1998),
but it too could not be satisfactorily explained (Wijers \& Galama 1999)
in a manner consistent with all of the other data on that burst.
In the case of \grb, additional IR data which do not show such a peak
have been reported by Garnavich et al. (2000).  In particular,
they find $K = 16.54 \pm 0.07$ on Dec. 18.25, which corresponds to
about 45\% more flux than our own nearly simultaneous 
measurement ($K = 16.89 \pm 0.17$ on Dec. 18.18) when extrapolated to
the same time.  If our own $K$-band point is in error, such a correction
would essential restore a power-law form to our optical and IR photometry,
so we suspect that this may be the case.

An additional constraint on the spectral slope and
extinction can be obtained in a weakly model-dependent
way by comparing the extrapolated
optical spectrum to the simultaneously measured X-ray flux.
In the X-ray, five measurements
were made between 1 and 40 hr after the burst, starting with the
$RXTE$ All-Sky Monitor (ASM, Corbet \& Smith 1999), continuing with
the $RXTE$ PCA (Takeshima et al. 1999), and 
ending with the {\it Chandra} High-Energy Transmission Grating
(HETG, Piro et al. 1999). 
All of these X-ray fluxes can be fitted by a power-law
temporal decay of index $\alpha_x = -1.616 \pm 0.067$,
as noted by the above authors (see Fig. 5).
This decay was used to determine
the X-ray flux on December 18.34 shown in Figure~6.
It is evident that
the smaller value of the extinction, $E(B-V) = 0.40$~mag,
is probably an underestimate, and that the larger value is possibly
an overestimate unless the spectrum steepens from $\beta = -0.5$ in the
optical to at least $\beta = -1$ from the ultraviolet through X-ray.
Allowing these extreme limits for Galactic extinction, the slope between
the $R$ band and the 2--10~keV X-rays can be characterized as
$\beta_{\rm ox} = -0.81 \pm 0.08$, although a broken power law
as illustrated in Figure~6 is allowed since the X-ray spectrum
itself has $\beta_{\rm x} \approx -1.1$ (Takeshima et al. 1999).

At the beginning of our optical monitoring, 10.8~hr after
the burst, there was also a simultaneous X-ray observation
(Takeshima et al. 1999).  At that earlier time the
$R$-band to X-ray spectrum of \grb\ can be described by
the slightly flatter index
$\beta_{\rm ox} = -0.74 \pm 0.05$, where
the error is again dominated by the uncertainty
in optical extinction suggested above.

\section{Geometry and Environment}

The early-time decay rate and broad-band spectral energy distribution 
of \grb\ from optical through X-ray are consistent with the standard theory
of adiabatic evolution in a uniformly dense medium (e.g.,
Sari, Piran, \& Narayan 1998).  In such a model,
a decaying synchrotron spectrum follows the form
$F(\nu,t) \propto \nu^{\beta}t^{\alpha}$ with $\alpha = (3/2)\beta =
-3(p-1)/4$ in the regime where $\nu < \nu_c$.  Here $p$ is the index
of the power-law electron energy distribution, and $\nu_c$ is the
``cooling frequency'' at which the electron energy loss time scale
is equal to the age of the shock.
At frequencies $\nu > \nu_c$,
the power law steepens by 1/2 to $\beta = -p/2$ because of synchrotron losses,
and $\alpha$ decreases by 1/4 to $-(3p-2)/4$.
At a time of
10.8~hr the optical-to-X-ray spectrum of \grb\ can be described as
$\beta_{\rm ox} = -0.74 \pm 0.05$.
At the same time the optical decay rate must follow
$\alpha_{\rm o} \geq -1.22$, depending upon whether
a simple power law or a dual power-law function
is fitted, which
is therefore consistent with $(3/2)\beta_{\rm ox}$.
Since in the X-ray $\beta_{\rm x} \approx -1.1$ (Takeshima et al. 1999),
the overall spectrum is consistent
with $\nu_c$ falling between the optical and X-ray at times between
10.8 and 40~hr, and $p \approx 2.4$.
Extrapolation of the X-ray spectrum to the
extension of the optical points shows that
$\nu_c \geq 1.2 \times 10^{16}$~Hz at 40~hr (Fig. 6).
Only the X-ray decay rate, $\alpha_{\rm x} \approx -1.6$
(Takeshima et~al. 1999), is slightly
discrepant from the predicted value which should be $\geq -1.47$.

The gradual steepening of the $R$-band decay is in accord with
some models of jet-like afterglows.  Numerous authors have discussed 
the possible effects of collimation.  At first, 
analytic arguments indicated that a steepening of the light curve
is expected after the edge of a jet is seen when
it slows to a Lorentz factor $\Gamma < \theta_0^{-1}$,
where $\theta_0$ is the initial opening angle of the jet
(Panaitescu et al. 1998).  At the same time,
or soon thereafter, the jet would begin to spread
(Rhoads 1999; Sari et al. 1999),
resulting in an asymptotic decay rate $\alpha_2 = -p$
and spectral shape which is constant in time.  Other authors
have shown through numerical simulations that such a transition,
if visible at all, is not very sharp.
The $\Gamma < \theta_0^{-1}$
transition produces at best a gradual transition to $\alpha = -2.0$
extended over two orders of magnitude in time (Moderski et al. 2000).
Kumar \& Panaitescu (2000) found that in a uniform
density medium the decay index steepens by $\sim 0.7$ over a factor
of 10 in time.  Both of these predictions are consistent with the
behavior of \grb, for which $\alpha$ in the range --1.5 to --2.1
describes the decay from 2 to $\sim 30$ days.  It has also been shown
that breaks are expected to result from the later transition
of a jet to the non-relativistic regime even when
they don't occur earlier in the relativistic phase (Huang et al.
1999,2000).  While the possible causes of temporal breaks
in afterglow decays are still uncertain, it is generally
agreed that when breaks are seen, collimated jets are 
likely to be responsible.

If the steepening in the optical decay of \grb\ to $\alpha \leq -1.5$
occurred at 1~d or later, then we need to explain why a steeper X-ray
decay  ($\alpha_x = -1.6$)
was observed at earlier times.  In the context of the 
$\Gamma < \theta_0^{-1}$
transition, this might be understood in
terms of the evolution of a layered jet, in which the higher-energy
emission is concentrated in a narrower core which began
to spread earlier.

The jet
theory is most consistent with the observations of \grb, while
an alternative interpretation of the observed steepening
as the passage of the cooling frequency $\nu_c$ through the
optical band is less plausible for three reasons.  First, the
expected decay exponent in the cooling regime is only $-(1.3-1.4)$,
which is not steep enough to account for the observations
at late times.  Second, $\nu_c$ declines as $t^{-1/2}$ in a
spherical afterglow.  Since 
$\nu_c \geq 1.2 \times 10^{16}$~Hz is observed at 40~hr,
we would expect $\nu_c \geq 9 \times 10^{15}$~Hz at 3~d,
which is still in the extreme ultraviolet.  Third, the decay
rate in the $I$ band through day~3 is at least as steep as
it is in the $R$ band.

Another class of models provides a better fit
to several of the GRB afterglows, but it is less compatible
with \grb.  This is the wind
interaction (Chevalier \& Li 1999,2000),
in which the afterglow develops in a nonuniform medium
of density $n \propto r^{-2}$ as appropriate for a pre-existing
stellar wind from a massive stellar progenitor.  In the wind model
$\alpha = (3\beta-1)/2 =
-(3p-1)/4$ for $\nu < \nu_c$, and the same evolution as the
constant density case applies for $\nu > \nu_c$.  If we make the 
plausible assumption that $p > 2$, then $\alpha < -1.25$ in
a wind environment, and the slow initial decay of \grb\ with
$\alpha_{\rm o} \geq -1.22$ is difficult to accommodate.
Alternatively, if we hypothesize that $\nu > \nu_c$
at the time of the earliest observations of the afterglow,
then the spectral index $\beta_{\rm ox} = -0.74$ is too flat to
meet the requirement for $\beta = -p/2$ in this cooling regime,
and the observed
spectral and temporal steeping in the X-rays is unexplained.
In wind models, the X-ray decay should not be steeper than the optical
decay, which is in contradiction to the observations of \grb.
We conclude that afterglow of \grb\ does not show a stellar wind
interaction, but behaves like a jet in a uniform ISM.

\section{Energetics and Origin}

Under the assumption that a jet-like GRB is collimated into the
same solid angle as its early afterglow, Sari et al. (1999)
argue that the cross-over time $t_*$ in the afterglow light curve
can be related to the $\gamma$-ray energy $E$ via
$$t_*\ \approx\ 6.2\,(E_{52}/n)^{1/3}\,(\theta_0/0.1)^{8/3}\,(1+z)
\ \ {\rm hr},
\eqno(2)$$
where $E_{52}$ is the apparent (isotropic) energy in units of
$10^{52}$~ergs, $n$ is the ISM density
in cm$^{-3}$, and $\theta_0$ is the half opening angle of the jet.
The factor $(1+z)$ is required if $t_*$ is in the observer's frame.
For the case of \grb, with $t_* \approx 2$~d and $E \approx 6.7 \times 10^{53}$~ergs, we infer that $\theta_0 \approx 6^{\circ}$,
and that the energy is reduced by a factor of $\approx 200$ to
$3.2 \times 10^{51}$~ergs, within the range of compact-object coalescence.
Even if $t_* \approx 5$~d, the energy reduction is still a factor of 100.
However, the assignment of the $\gamma$-ray energy to $E$ in this analysis
is not an obvious choice.  The energy powering the afterglow
expansion could be either more or less than the observed $\gamma$-rays.
As an alternative, we can estimate the observed energy in the afterglow itself,
which is dominated by the X-rays at early times.  If we integrate the
observed (2--10~keV) X-ray flux back to a time of 600~s, when its flux
would have been $\approx 1.1\,$mJy, comparable to the radio peak,
we get $\approx 2.1 \times 10^{-5}$~ergs~cm$^{-2}$, or almost 10\%
of the $\gamma$-ray burst fluence in less than one decade of frequency.
Thus, it seems that we cannot be far from wrong in using either the burst
or the afterglow energy in Equation (2).  In particular, we are probably
not underestimating the opening angle of the jet.  Also, because of the
extreme energy, it is unlikely that the jet has become non-relativistic
during the times considered here.  For this to have occurred, densities
in excess of $10^4$~cm$^{-3}$ would be required, and there is little
evidence for the excess extinction that would be expected from
such an environment.

\grb\ is the third good example
of a jet-like afterglow following GRB~990123 (Kulkarni et al. 1999)
and GRB~990510 (Harrison et al. 1999; Stanek et al. 1999;
Beuermann et al. 1999),
supporting a trend in which the apparently
most energetic $\gamma$-ray events have the narrowest collimation
{\it and} a uniform ISM environment.  
[The only other event with a demonstrated isotropic energy $> 10^{53}$~ergs
was GRB~971214 at $z = 3.42$ (Kulkarni et al. 1998),
but its afterglow was not
well characterized because it was both faint and reddened.]
Chevalier \& Li (2000) classified afterglows into two types
according to whether their evolution best matches an ISM (constant
density) or stellar wind ($n \propto r^{-2}$) environment.
Only GRB~990123 and GRB~990510 definitely fell in the ISM category,
leading Chevalier \& Li to speculate that these were compact-object
mergers. They also noted an absence of evidence
for supernovae associated with these jet-like afterglows, although
supernovae may be difficult to see in these hosts at $z > 1$
(Bloom et al. 1999).
In the case of GRB~990510, no host galaxy has been found to a limiting
magnitude of $V > 28$ (Fruchter et al. 1999; Beuermann et al. 1999).
Although host galaxies as faint as this are not unexpected
(Hogg \& Fruchter 1999),
the possibility is at least allowed that GRB~990510 occurred
outside its parent galaxy.  A massive star is expected to explode
close to its birth site, whereas an evolved compact binary may 
or may not escape its parent galaxy.  If
the mechanism of collimation, probably magnetic in nature,
can be at least as effective in compact 
binary mergers as it is in the collapse of
a single massive star (MacFadyen \& Woosley 1999),
then the merger is a plausible origin of jet-like bursts.
However, this would certainly contradict the prevailing
theory (e.g., Fryer, Woosley, \& Hartmann 1999)
that massive stars are the progenitors of the
long-duration GRBs which comprise all of the ones 
that have been localized, while compact
binary mergers are responsible for the as-yet
unidentifed sources of the short-duration GRBs.

\acknowledgments

We thank Sebastiano Novati for his help with the initial observations
at MDM Observatory.
 
\clearpage

%
%

\clearpage

\begin{figure}
\plotone{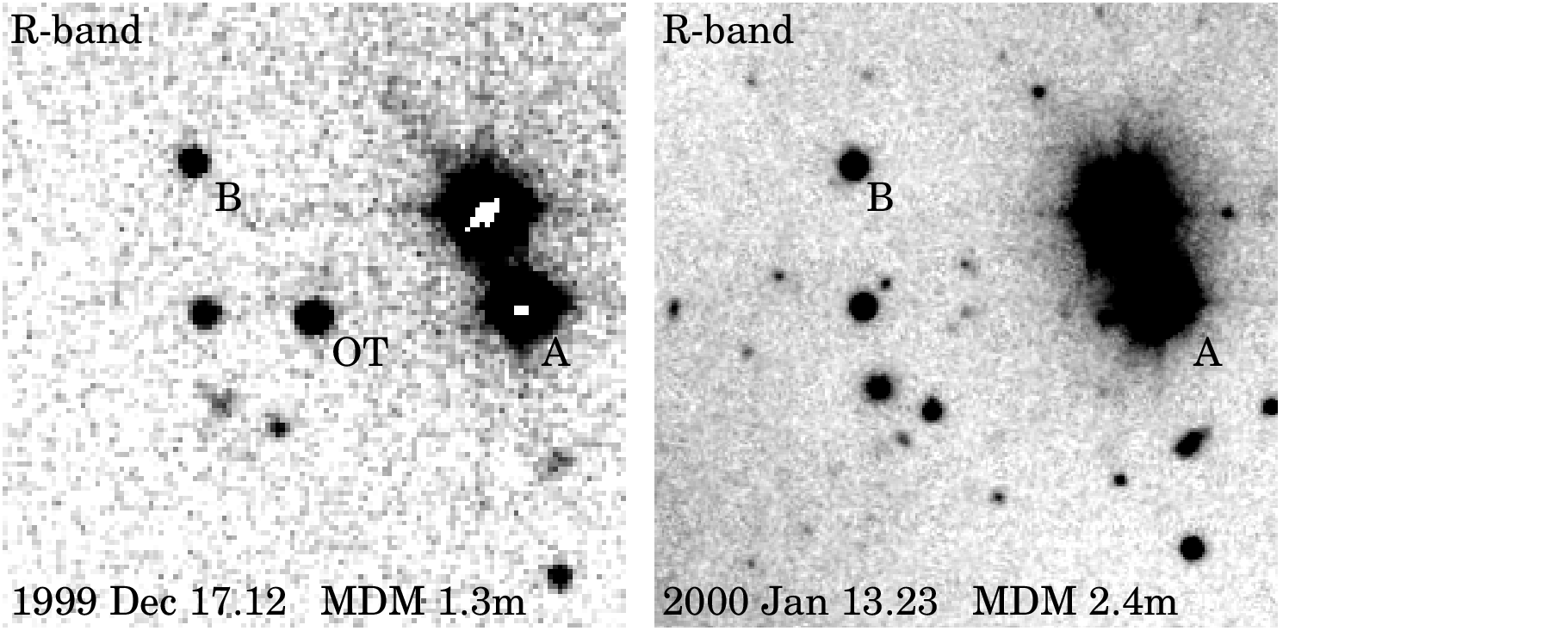}
\caption{Images of the \grb\
optical transient at discovery 10.8~hr after the burst when $R = 18.5$,
and after 4 weeks when $R = 24.0$ and a galaxy dominates.  Approximately
$1^{\prime}\times1^{\prime}$ sections of the images are displayed.
The position of the OT is
(J2000) $05^{\rm h}09^{\rm m}31.\!^{\rm s}29,
+11^{\circ}17^{\prime}07.\!^{\prime\prime}4$.
Comparison stars A and B named by Jha et al. (1999)
and used in various GCN circulars
are indicated.  North is up, and east is to the left. \label{fig1}}
\end{figure}

\begin{figure}
\plotone{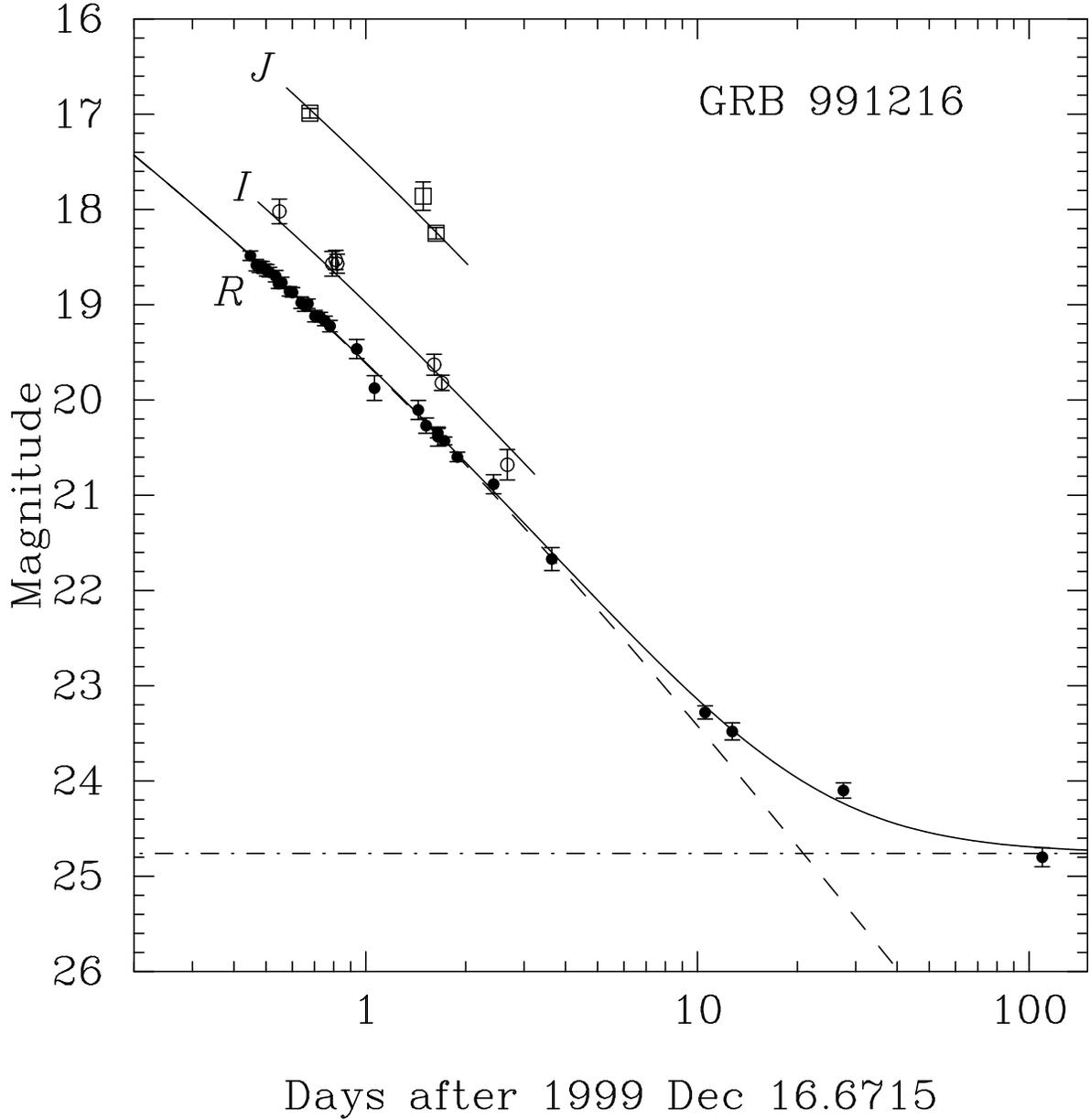}
 \caption{Light curves of \grb\ in $RIJ$ colors.
The data are taken from Table~1.
The {\it solid line} is a
model of the $R$-band decay consisting of a dual power law ({\it dashed line})
plus constant ({\it 
dot-dashed line})
as parameterized by Equation (1).  For illustrative
purposes only, $\alpha_1 = -1.0$ was fixed,
which results in best-fitted parameter values $\alpha_2 = -1.8$,
$t_* = 1.2$~d, and $R_0 = 24.76$.  The latter is the constant
contribution of the portion of the
host or intervening galaxy which contaminates the OT photometry.
Solid lines fitted to the $I$ and $J$ data
have the same values of all parameters,
but are offset by a constant. \label{fig2}}
\end{figure}

\begin{figure}
\plotone{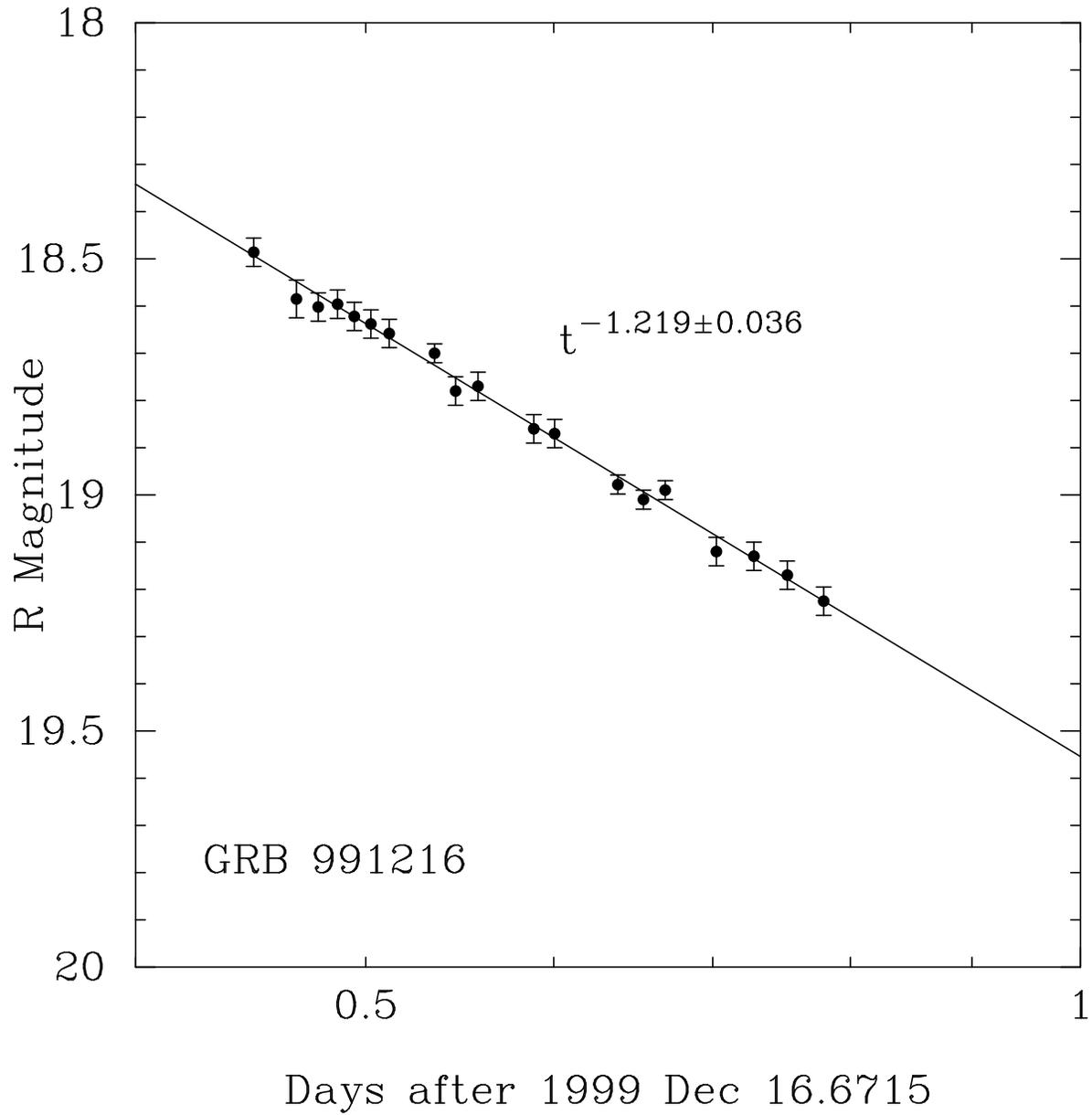}
\caption{Fit of a power law to the first night of
data from the MDM 1.3m telescope.  Only statistical errors are
employed in this figure, since all of the points are from the
same instrument. \label{fig3}}
\end{figure}

\begin{figure}
\plotone{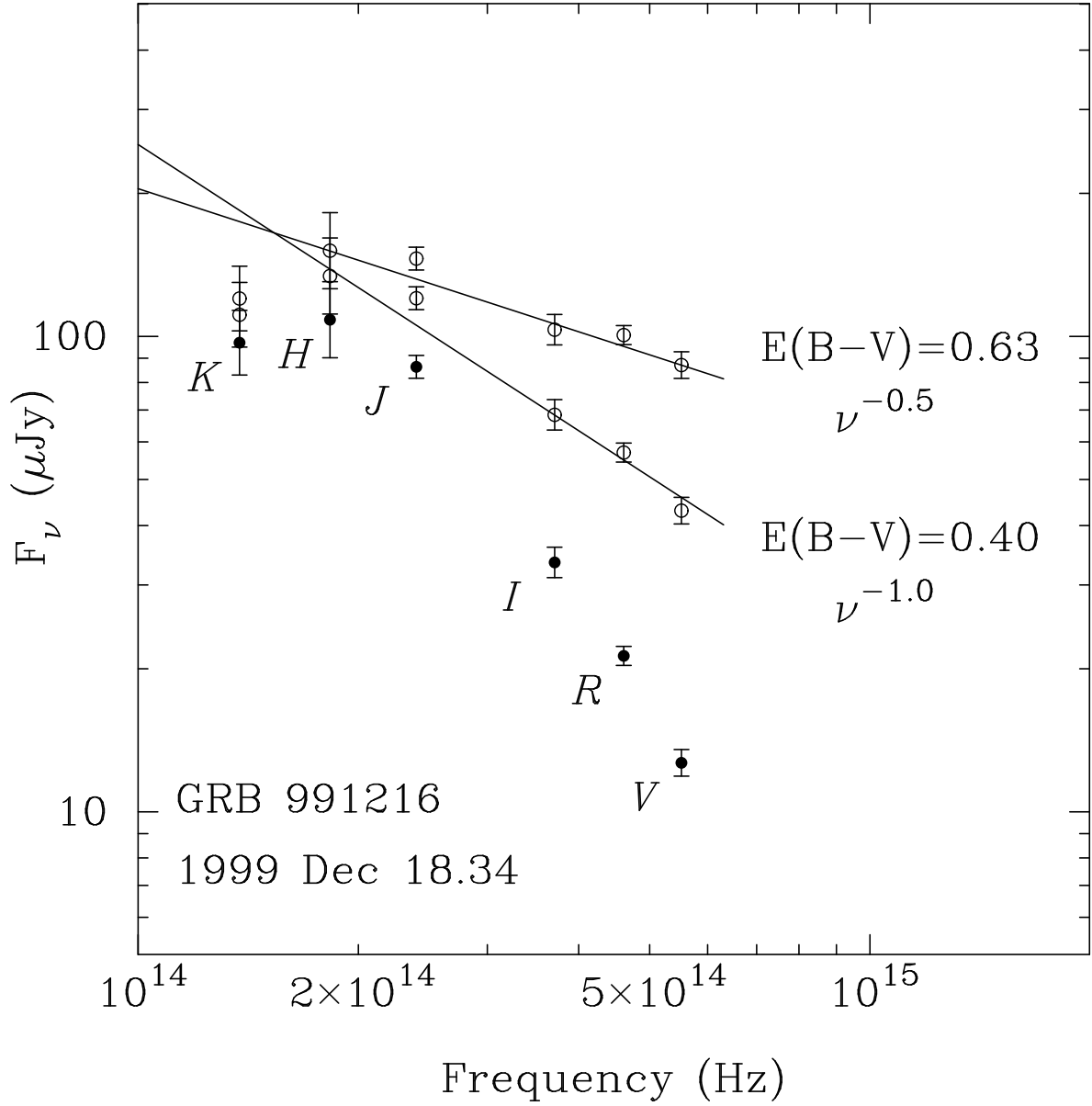}
\caption{Synthetic spectrum of \grb\ 40~hr after
the burst, constructed from the data in Table~1 ({\it filled circles}).
Two different estimates
of the Galactic extinction, as described in the text, are used to
deredden the fluxes ({\it open circles}), and they suggest very
different power-law slopes ({\it solid lines}).
\label{fig4}}
\end{figure}

\begin{figure}
\plotone{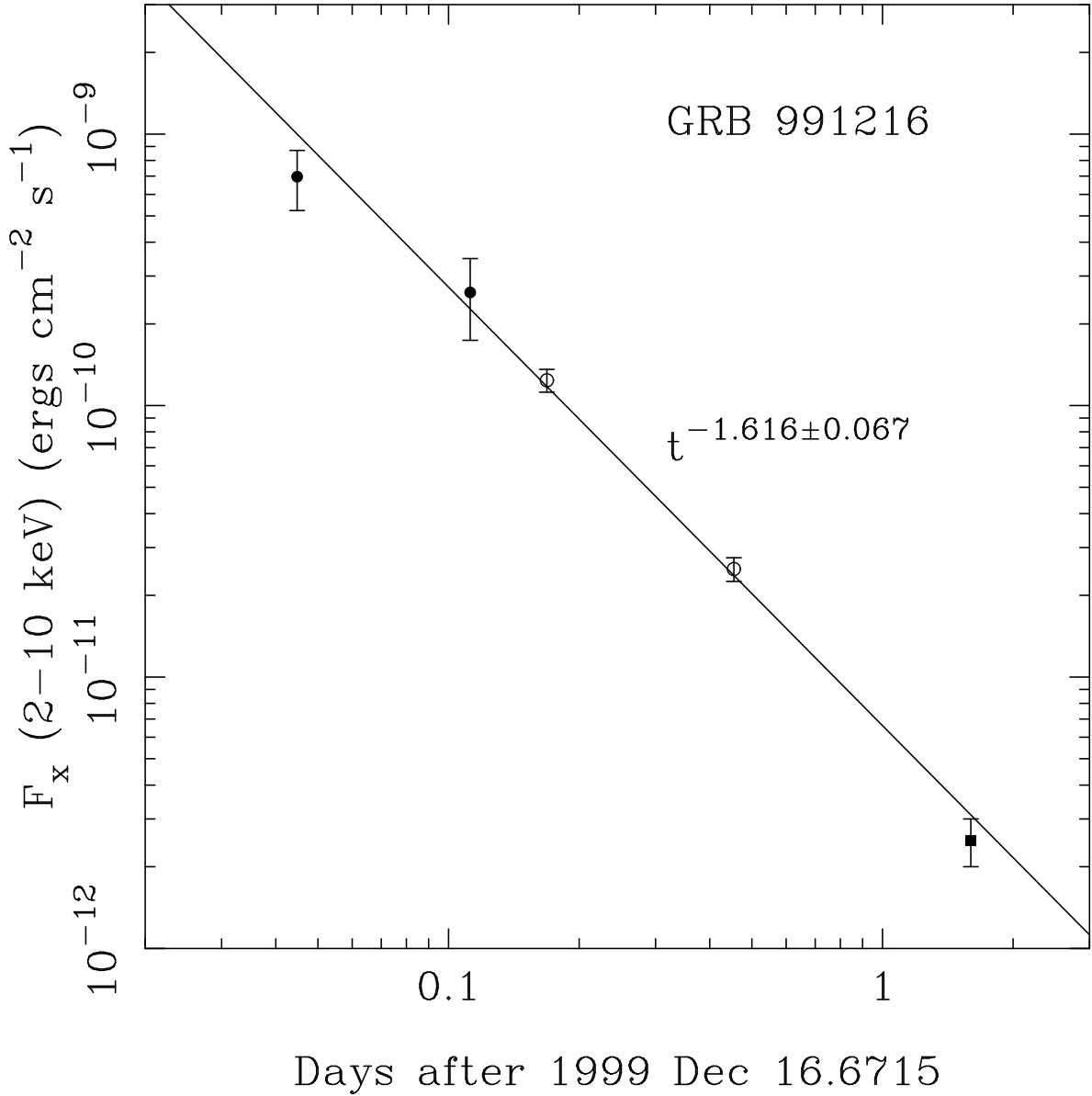}
\caption{X-ray decay of \grb\ as measured by the
$RXTE$ ASM ({\it filled circles}, Corbet \& Smith 1999),
the $RXTE$ PCA
({\it open circles}, Takeshima et al. 1999), and 
the {\it Chandra} HETG
({\it filled square}, Piro et al. 1999).  We increased the 
very small error bars quoted by Takeshima et al. to $\pm 10\%$ 
in order to allow for possible systematic differences in calibration
among the instruments.\label{fig5}} 
\end{figure}

\begin{figure}
\plotone{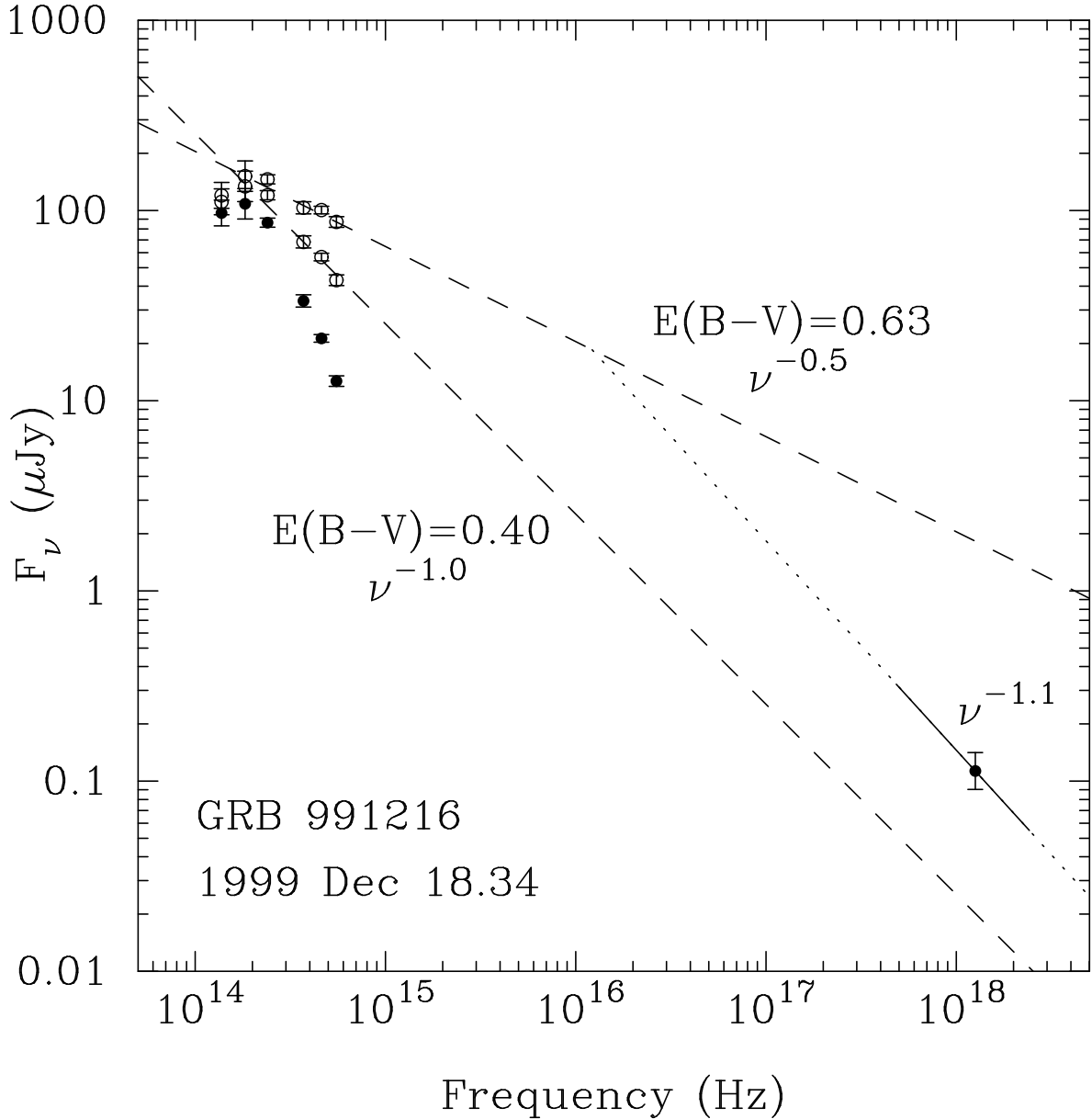}
\caption{A broad-band view of the spectrum of
\grb\ 40~hr after the burst, including the X-ray flux
observed by {\it Chandra} as described by Piro et al. (1999).
As in Figure~4, two different estimates of
the Galactic extinction are used to deredden the optical/IR fluxes.
It is likely that the true reddening and optical-to-X-ray
spectral slope lie in between the extreme values shown 
({\it dashed lines}).  The observed (2--10~keV)
X-ray spectrum is indicated ({\it solid line}),
as is its extrapolation ({\it dotted line}).\label{fig6}}
\end{figure}

\clearpage
 
\begin{deluxetable}{llcrr}
\footnotesize
\tablecaption{Optical and IR Photometry of GRB 991216. \label{tbl-1}}
\tablewidth{0pt}
\tablehead{
\colhead{Date (UT)} & \colhead{Telescope}   & \colhead{Filter}   & 
\colhead{Magnitude} &  \colhead{Reference}} 
\startdata
1999 Dec 18.341  & Lowell 1.8m & $V$ & $21.13 \pm 0.07$ & 1 \nl
1999 Dec 17.12   & MDM 1.3m    & $R$ & $18.49 \pm 0.05$ & 1 \nl
1999 Dec 17.139  & MDM 1.3m    & $R$ & $18.59 \pm 0.06$ & 1 \nl
1999 Dec 17.149  & MDM 1.3m    & $R$ & $18.60 \pm 0.06$ & 1 \nl
1999 Dec 17.153  & Lowell 1.8m & $R$ & $18.52 \pm 0.08$ & 1 \nl
1999 Dec 17.157  & Lowell 1.8m & $R$ & $18.64 \pm 0.06$ & 1 \nl
1999 Dec 17.158  & MDM 1.3m    & $R$ & $18.60 \pm 0.05$ & 1 \nl
1999 Dec 17.166  & MDM 1.3m    & $R$ & $18.62 \pm 0.05$ & 1 \nl
1999 Dec 17.174  & MDM 1.3m    & $R$ & $18.64 \pm 0.06$ & 1 \nl
1999 Dec 17.183  & MDM 1.3m    & $R$ & $18.66 \pm 0.05$ & 1 \nl
1999 Dec 17.206  & MDM 1.3m    & $R$ & $18.70 \pm 0.06$ & 1 \nl
1999 Dec 17.217  & MDM 1.3m    & $R$ & $18.78 \pm 0.05$ & 1 \nl
1999 Dec 17.229  & MDM 1.3m    & $R$ & $18.77 \pm 0.06$ & 1 \nl
1999 Dec 17.260  & MDM 1.3m    & $R$ & $18.86 \pm 0.05$ & 1 \nl
1999 Dec 17.272  & MDM 1.3m    & $R$ & $18.87 \pm 0.05$ & 1 \nl
1999 Dec 17.310  & MDM 1.3m    & $R$ & $18.98 \pm 0.06$ & 1 \nl
1999 Dec 17.326  & MDM 1.3m    & $R$ & $19.01 \pm 0.06$ & 1 \nl
1999 Dec 17.340  & MDM 1.3m    & $R$ & $18.99 \pm 0.05$ & 1 \nl
1999 Dec 17.374  & MDM 1.3m    & $R$ & $19.12 \pm 0.06$ & 1 \nl
1999 Dec 17.40   & MDM 1.3m    & $R$ & $19.13 \pm 0.05$ & 1 \nl
1999 Dec 17.424  & MDM 1.3m    & $R$ & $19.17 \pm 0.05$ & 1 \nl
1999 Dec 17.451  & MDM 1.3m    & $R$ & $19.23 \pm 0.06$ & 1 \nl
1999 Dec 17.61   & Hawaii 2.2m & $R$ & $19.47 \pm 0.10$ & 2 \nl
1999 Dec 17.733  & Wise 1m     & $R$ & $19.88 \pm 0.13$ & 3 \nl
1999 Dec 18.11   & Danish 1.5m & $R$ & $20.11 \pm 0.10$ & 4 \nl
1999 Dec 18.191  & Lowell 1.8m & $R$ & $20.27 \pm 0.08$ & 1 \nl
1999 Dec 18.32   & Danish 1.5m & $R$ & $20.39 \pm 0.10$ & 4 \nl
1999 Dec 18.398  & Palomar 5m  & $R$ & $20.43 \pm 0.04$ & 1 \nl
1999 Dec 18.32   & Hawaii 2.2m & $R$ & $20.35 \pm 0.05$ & 5 \nl
1999 Dec 18.56   & Hawaii 2.2m & $R$ & $20.60 \pm 0.05$ & 5 \nl
1999 Dec 19.10   & NOT 2.5m    & $R$ & $20.89 \pm 0.10$ & 4 \nl
1999 Dec 20.31   & Lowell 1.8m & $R$ & $21.67 \pm 0.12$ & 1 \nl
1999 Dec 27.217  & MDM 2.4m    & $R$ & $23.28 \pm 0.07$ & 1 \nl
1999 Dec 29.405  & Keck II     & $R$ & $23.48 \pm 0.09$ & 1,6 \nl
2000 Jan 13.232  & MDM 2.4m    & $R$ & $24.10 \pm 0.08$ & 1 \nl
2000 Apr 4.23    & Keck II     & $R$ & $24.80 \pm 0.10$ & 1 \nl
1999 Dec 17.220  & Palomar 5m  & $I$ & $18.02 \pm 0.13$ & 1 \nl
1999 Dec 17.463  & Palomar 5m  & $I$ & $18.57 \pm 0.13$ & 1 \nl
1999 Dec 17.483  & MDM 1.3m    & $I$ & $18.53 \pm 0.10$ & 1 \nl
1999 Dec 17.491  & MDM 1.3m    & $I$ & $18.57 \pm 0.10$ & 1 \nl
1999 Dec 18.278  & Palomar 5m  & $I$ & $19.63 \pm 0.11$ & 1 \nl
1999 Dec 18.366  & Lowell 1.8m & $I$ & $19.82 \pm 0.08$ & 1 \nl
1999 Dec 19.342  & HET         & $I$ & $20.68 \pm 0.16$ & 1 \nl
1999 Dec 17.35   & FLWO 1.2m   & $J$ & $16.99 \pm 0.05$ & 5 \nl
1999 Dec 18.16   & MDM 2.4m    & $J$ & $17.86 \pm 0.15$ & 1 \nl
1999 Dec 18.30   & FLWO 1.2m   & $J$ & $18.25 \pm 0.06$ & 5 \nl
1999 Dec 18.21   & MDM 2.4m    & $H$ & $17.34 \pm 0.20$ & 1 \nl
1999 Dec 18.18   & MDM 2.4m    & $K$ & $16.89 \pm 0.17$ & 1 \nl
\enddata
\tablenotetext{}{References.--(1) this paper; (2) Jha et al. (1999);
(3) Leibowitz (1999); (4) Jensen et al. (1999); (5) Garnavich et al. (1999);
(6) Djorgovski et al. (1999).} 
\end{deluxetable}

\end{document}